\documentstyle[12pt]{article}
\newcommand{\be}{\begin{equation}}
\newcommand{\ee}{\end{equation}}
\newcommand{\bea}{\begin{eqnarray}}
\newcommand{\eea}{\end{eqnarray}}
\newcommand{\beax}{\begin{eqnarray*}}
\newcommand{\eeax}{\end{eqnarray*}}
\newcommand{\no}{\nonumber\\}

\begin{document}

PACS 11.10.Gh, 12.15.Cc, 64.60.Ah
\hfill  CTP \#2477

\hfill  October, 1995

\vskip 0.5in                         
\centerline{\large Non-Trivial Directions for Scalar Fields}
\vskip 0.25in  
\centerline{Kenneth Halpern and Kerson Huang}
\vskip 0.25in
\centerline{Center for Theoretical Physics, Laboratory for 
Nuclear Science}
\centerline{       and Department of Physics}
\centerline{   Massachusetts Institute of Technology}
\centerline{   Cambridge, MA 02139}
\vskip 0.5in
\centerline{                ABSTRACT}
\bigskip
We study the eigenvectors of the renormalization-group matrix for
scalar fields at the Gaussian fixed point, and find that
that there exist ``relevant'' directions in parameter space.
They correspond to theories with exponential potentials that are nontrivial
and asymptotically free. All other potentials, including polynomial potentials,
are ``irrelevant,'' and lead to trivial theories. Away from the 
Gaussian fixed point, renormalization does not induce
derivative couplings, but it generates non-local interactions.

\section{Introduction and Summary}
In a previous note [1], we discussed the renormalization group (RG) 
for scalar field theories, and reported RG trajectories near
the Gaussian fixed point along which the scalar 
theory is nontrivial and asymptotically free. In this paper, we give the 
details, including a critical analysis of the calculations.
In particular, we address the question of whether renormalization generates 
interactions not originally present in the Lagrangian.

To address the question of closure under RG, we start with the most general
action conceivable for a real scalar field $\phi(x)$ in $d$ space-time 
dimensions. Eventually, we focus our attention on a theory with local 
non-derivative couplings, whose Euclidean action is given by
\bea
A[\phi]
&=& \int d^dx \left[ {1\over 2}(\partial\phi)^2 + U(\phi^2) \right]\no
U(\phi^2)&=& g_2 \phi^2 + g_4 \phi^4 +\cdots
\label{e:action}
\eea
The potential $U(\phi^2)$ is arbitrary, and not necessarily 
polynomial. For simplicity we work with a one-component 
real field here; but extension to a multicomponent field with $O(N)$ symmetry
is straightforward, and we shall quote results for that case. 
There is a high-momentum cutoff $\Lambda$. To make calculations 
feasible, we use a sharp cutoff, which also proves to be 
a limitation, for it leads to ambiguous non-local interactions.
We only report results that are believed to be independent
of the cutoff function.

Scalar fields are used in the Higgs sector of the
standard model, where it is customary to assume that $U(\phi^2)$ is  quartic 
in $\phi$. It turns out that such a choice leads to ``triviality,'' 
in that the renormalized value of $g_4$ vanishes in the limit 
$\Lambda\to\infty$, and one is left with a free field. This 
startling result was implicit in the work of 
Larkin and Khumel'nitskii [2], and demonstrated by Wilson [3]. It 
has been verified in a number of independent Monte-Carlo simulations [4]-[8]. 
There are proposals on how to deal with this awkward situation:

(a) Physical quantities are insensitive to the value of the cutoff, because
the approach to the free-field limit proceeds with logarithmic slowness [2].
Thus, one can keep the cutoff finite, as a parameter of the model. 
Considerations of self-consistency [9] impose an upper bound,
estimated to be 600 GeV [8], on the Higgs mass.

(b) Even in the free-field limit, the theory is not entirely trivial. The 
field can have a non-vanishing vacuum expectation, as shown in Monte Carlo 
simulations [6]. Thus it can still be used as a phenomenological method
to generate particle masses.

These alternatives are not completely satisfactory, for they do not
take the field theory seriously. The purpose of this paper is to show that 
escape routes do exist in the framework of renormalized quantum field theory.
In the rest of this section, we describe our approach to the problem, 
and summarize the results.

Common belief holds that only $\phi^4$ theories are renormalizable, 
in the sense that higher powers in the potential will give 
Feynman graphs requiring an infinite number of subtraction constants. 
This is true if the higher coupling constants, which 
generally have dimensions, set independent scales. From a physical 
standpoint, however, these scales contain information about the system
at momenta higher than $\Lambda$, of which nothing is supposedly
known. Accordingly, we shall assume that $\Lambda$ is the only intrinsic 
scale in the problem.  This means that all coupling constants 
should be scaled by appropriate powers of the cutoff:
\be
g_\alpha=u_\alpha\Lambda^{\alpha+d-\alpha d/2}
\ee
where the $u_\alpha$ are dimensionless parameters. These factors of $\Lambda$ 
supply extra convergence to Feynman graphs, rendering them
renormalizable in the usual sense [10]. It can be shown that
the S-matrix of the theory in $d=4$ is the same
as that of an effective $\phi^4$ theory, whose effective coupling is
a function of the $u_\alpha$ [1]. However, the RG behavior of the effective 
coupling is not the same as that of a $\phi^4$ theory, for it depends on 
the RG flow of the $u_\alpha$, which can only be obtained from the 
original theory. 

Renormalization relates the coupling constants at
different momentum scales. In Wilson's formulation [3], the relation 
is found through a RG transformation that represents a 
coarse-graining process,  eliminating the degrees of freedom 
with momenta between $\Lambda$ and $\Lambda/b$, and effectively lowering 
the cutoff by a factor $b$. The new action should have
the same form as the old, except that the ``bare'' couplings 
$u_\alpha$ are replaced by the  ``renormalized'' ones $u'_\alpha$. 
Making an infinitesimal RG transformation in the neighborhood 
of $b=1$ yields differential equations for $u_\alpha$, the RG equations. 
They generate RG trajectories in the parameter space 
spanned by the $u_\alpha$. The flow along a trajectory always proceeds
in the coarse-graining direction, {\it i.e.}, direction of increasing 
length scale. If $A$ and $B$ are two points on
a trajectory, with the sense of flow from $A$ to $B$, then $A$ 
corresponds to a bare system, and $B$ a renormalized system.

It should be emphasized that the cutoff $\Lambda$ does not appear in $A[\phi]$
explicitly, for we can set $\Lambda=1$ by choosing appropriate units. 
Its value is reflected solely in the values of the coupling 
constants $u_\alpha$. Thus, the RG equations give the tangent vector to
a trajectory at an arbitrary point.

The actual value of the cutoff can be deduced only by computing some 
physical quantity, such as the correlation length. Thus, the only way to 
approach the limit $\Lambda\to\infty$ is to go to some point 
in the parameter space at which 
the correlation length is infinite. Since the length scale increases
under an RG transformation, such a point must be  a fixed point, where
the system is invariant under RG transformations.

If a trajectory flows into a fixed point (in the coarse-graining direction,)
then, to systems lying on that trajectory, the fixed point is infrared (IR),
representing the low-energy limit of the theory. 
If a trajectory flows out of a fixed point,
then to systems on this trajectory the fixed point is ultraviolet (UV),
corresponding to the high-energy limit of the theory.

Although we are free to choose a bare action, the renormalized action is 
determined by the RG transformation, and is not under our control. 
For example, if we start with a $\phi^4$ 
theory at some value of the cutoff, an RG transformation may generate 
$\phi^6$ and other couplings. Only at a fixed point are the
the couplings determined. When we 
approach a fixed point along a trajectory, in the coarse-graining sense, 
some couplings 
not destined to be in the fixed-point action will tend to zero, 
and these are called ``irrelevant'' couplings. Conversely, when we go away 
from a fixed point in a coarse-graining sense, some
couplings that were infinitesimally small will grow,
and these are termed ``relevant.'' Triviality comes from the existence of
a IR fixed point at zero couplings, the Gaussian 
fixed point. By examining all possible trajectories in the 
neighborhood of the Gaussian fixed point, we find 
that, although the fixed point is IR in theories with polynomial
potentials, it is UV to a class of potentials with exponential 
asymptotic behavior.

To insure that the parameter space is closed under RG transformations,
we have to consider an arbitrary action, which should include
derivative couplings as well as non-local interactions. 
A derivative coupling refers to terms containing a derivative of the field
not of the form of the kinetic term $\int d^dx (\partial\phi)^2$,
as for example
\be
\int d^dx (\partial^2\phi)^2
\ee
A non-local term involves fields or derivatives at different space-time
points, as for example
\be
\int d^dx d^dy \phi(x)K(x-y)\phi(y)
\ee
Actually, the action with a momentum cutoff is non-local
within a spatial distance of order $\Lambda^{-1}$. By ``non-local terms,'' 
we specifically refer to those for which the range of non-locality is
large compared to $\Lambda^{-1}$.

The exact RG equations for the most general case have been obtained by 
Wegner and Houghton [11], and we shall review the derivation. 
This remarkable calculation is made possible by the simplicity of
the sharp cutoff. The equations
show that RG transformations do not induce derivative couplings 
if none were present from the start. On the other hand, non-local terms are
always generated. Some of these have infinite range, being of the form 
$V^{-1}[\int d^dx \phi(x)]^2$, where $V$ is the space-time volume.
Though consistent with the fact that the action is $O(V)$, such a term is
indeterminate in the limit $V\to\infty$. The ambiguity can be ascribe to the
infinitesimal RG transformation made with a sharp momentum cutoff. 
It would disappear if  gentle 
cutoff functions were used, or if the momentum-shell integration had extended 
over a finite instead of an infinitesimal shell. Both of these alternatives,
however, make the problem intractable. 

Fortunately, the ambiguous non-local terms are second order in the
bare couplings. We can therefore neglect them in a linear approximation
about the the Gaussian fixed point, and the action 
(\ref{e:action}) becomes closed under RG in this approximation.
We study the eigenvalue problem based on the RG matrix, which should 
be insensitive to the form of the cutoff. It tell us about the characteristics
of various ``principal axes'' in parameter space at the origin.
Our main results are as follows:

(a) There exist trajectories flowing into the Gaussian fixed point, 
as well as flowing out of it. That is, the Gaussian fixed point is IR with 
respect to some trajectories, and UV with respect to others.

(b) For all theories with polynomial potentials, the Gaussian fixed 
point is IR. These theories are consequently trivial. A similar
result was obtained earlier by Hasenfratz and Hasenfratz [12].

(c) For a class of non-polynomial potentials, the Gaussian 
fixed point is UV.  For $d>2$, potentials in this class behave like
$U(\phi)\sim\exp[c(d-2)\phi^2]$ for large $\phi$, where $c$ is a constant.
Theories with such potentials are nontrivial and asymptotically free.
Some of the potentials exhibit spontaneous symmetry breaking.

In summary, we can say that in a sufficiently small
neighborhood of the Gaussian fixed point, conventional scalar 
theories with polynomial interactions are trivial, and that certain 
models with exponential potentials are non-trivial. 
For conventional potentials, the road to oblivion is clear and
inescapable, because with each RG step we are closer to the fixed point, and
the linear approximation improves. For the non-trivial models, 
on the other hand, the escape route is clouded, since RG steps tend to take us 
out of the linear region into unknown territory.

\section{Renormalization Procedure}

We shall begin with the most general scalar field theory, with arbitrary 
derivative and non-local couplings, and choose units such that
the cutoff momentum is unity:
\be
\Lambda=1
\ee
We enclose the system in a periodic hypercube of volume $V$, and define
the Fourier transform of the field by
\be
\phi_k=V^{-1/2}\int d^dx e^{-ik\cdot x} \phi(x)
\ee
with $\phi_k^\ast=\phi_{-k}$. Eventually, we take the limit $V\to\infty$,
in which the Fourier component is replaced by the continuum
version $\phi(k)=V^{-1/2}\phi_k$. For illustration, the action 
(\ref{e:action}) can be written as
\be
A[\phi]={1\over 2}\sum_{|k|<1} (k^2+r)\phi_k\phi_{-k} +
{u_4\over V}\sum_{|k_i|<1} \delta(k_1+\cdots+k_4)
\phi_{k_1}\cdots\phi_{k_4}+\cdots
\ee
where $\delta(k)$ is the Kronecker delta $\delta_{k0}$.

To generalize the action, all we have to do is to replace $u_\alpha$ by 
an arbitrary function $u_\alpha(k_1,\ldots,k_\alpha)$, which we abbreviate 
as $u_\alpha(k)$. Thus, our starting point is the action
\be
A[\phi] = \sum_{\alpha=2}^\infty V^{1-\alpha/2}\sum_{|k_i|<1}  
\delta(k) u_\alpha(k) \phi_{k_1}\cdots\phi_{k_\alpha}
\label{e:gaction}
\ee
where $\delta(k)$ is an abbreviation for $\delta(k_1+\cdots+k_\alpha)$.
Without loss of generality, we may assume that $u_\alpha(k)$ is a 
symmetric function of its arguments.
To fix the normalization of the field, we normalize $u_2(k_1,k_2)$
as follows:
\be
v(k)\equiv 2 u_2(k,-k) = k^2 + r + c_4 k^4 + c_6 k^6 +\cdots
\label{e:norm}
\ee
The generalized kinetic term is
\be
A_2[\phi]\equiv{1\over 2} \sum_{|k|<1} v(k) \phi_k\phi_{-k}=
{1\over 2} \int\limits_{|k|<1}{d^dk\over (2\pi)^d} v(k) \phi(k)\phi(-k)
\ee
from which we can see that $v(k)=v(-k)$ is the inverse propagator 
for Feynman graphs. 

Wilson's RG transformation [3] [13] is defined in terms of the partition 
function
\be
Z=\int D\phi e^{-A[\phi]}
\ee
The object is to eliminate the Fourier components with momentum magnitudes
between 1 and $1/b$, without changing the partition function. We decompose
the field into a ``slow'' part $S_k$ and a ``fast'' part $f_k$:
\be
\phi_{k}=S_{k} + f_{k}
\ee
where
\bea
S_k &=& 0\quad{\rm unless\ } |k|<1/b\no
f_k &=& 0\quad{\rm unless\ } 1/b \le |k| \le 1
\eea
Let us split off the kinetic term in the action by writing
\be
A[\phi]=A_2[\phi] + A_I[\phi]
\ee
where $A_I$ is the ``interaction'' part. Since $S_k f_{-k} =0$, as their 
domains do not overlap, $A_2[S+f]$ is additive:
\be
A_2[S+f]=A_2[S]+A_2[f]
\ee
We now write
\bea
Z&=&\int DS \int Df e^{-A_2[S]-A_2[f]-A_I[S+f]}\no
&=&N\int DS e^{-A_2[S]}
\left\langle e^{-A_I[S+f]}\right\rangle_f
\equiv N\int DS e^{-\tilde A[S]}
\eea
where $N$ is a constant, and 
$\left\langle O\right\rangle_f$ denotes averaging over $f$ with weight
$\exp\{ -A_2[f]\}$. The new action 
\be
\tilde A[S] \equiv A_2[S]
-\ln \left\langle e^{-A_I[S+f]}\right\rangle_f
\ee
contains only the slow fields, with the the cutoff lowered to $1/b$. 
Writing out the first few terms, we have
\be
\tilde A[S] = {1\over 2}\sum_{|k|<1/b} [z k^2 + r_1 + \cdots]
S_k S_{-k}
+\cdots
\label{e:atilde}
\ee
The parameters $z$, $r_1$, {\it etc.} are proportional $b^{-y}$, 
where $y$ is a characteristic index.

To make comparison with the original action, we must restore the cutoff to 1, 
and normalize the field according to the convention (\ref{e:norm}).
The cutoff can be restored by changing the momentum integration 
variable to
\be
k'=bk
\label{e:kprime}
\ee
To restore the normalization, we transform the field to
\be
\phi'_{k'} \equiv S_{k'/b} b^{-1-d/2-\eta/2}
\label{e:phiprime}
\ee
where $\eta$ is the index of $z$ in (\ref{e:atilde}), {\it i.e.}, 
$z=b^{-\eta}$. The partition function can now be put in the form
\be
Z = N\int D\phi' e^{-A'[\phi']}
\ee
where
\be
A'[\phi'] \equiv \tilde A[S]
\ee
The action $A'[\phi']$ should have the same form as $A[\phi]$ in 
(\ref{e:gaction}), except that  the bare coupling function
$u_\alpha(k)$  is replaced by  the renormalized coupling function 
$u'_\alpha(k')$, which is of course a function of $b$.

The RG transformation can be formulated in terms of Feynman graphs.
By expanding $\exp\{-A_I[S+f]\}$
in powers of $f$, we can obtain $\tilde A[S]$ as a sum 
of connected Feynman graphs, in which all external momenta are
``slow,'' while all internal momenta are ``fast.'' That is, an external
line is associated with $S_k$; an internal line
is associated with $f_k$, and gives the propagator $1/v(k)$
after functional integration weighted with $A_2[f]$.  A vertex represents
a momentum-dependent factor $u_\alpha(k)$. 

\section{Renormalization-Group Equations}

We shall carry out an infinitesimal RG transformation at the cutoff momentum. 
The fast momenta are contained in a shell $\sigma$ in momentum space:
\be
\sigma=\{k| e^{-t}<|k|<1\}
\ee
where we have put $b=e^t$. Calculating to first order in $t$ will 
yield equations for $du_\alpha/dt$, which are the RG equations. To this order,
all internal momenta in Feynman graphs are integrated over a shell 
of infinitesimal thickness $t$, just below the surface of the unit sphere. 
Each independent integration therefore yields $O(t)$. This circumstance leads 
to the following simplifications: 

(a) To first order, we need to keep only tree and one-loop graphs.

(b) A one-loop graph with two or more vertices must have two or more
propagators, and is superficially $O(t^2)$. But it is $O(t)$ when the 
total momentum of the external lines emerging from any one vertex is
zero. An equivalent statement is that all the internal lines should 
carry exactly the same loop momentum.

To show (b), consider the simple one-loop graph in Fig.1,
which is proportional to
\be
\int\limits_{k_1\in\sigma} d^dk_1\int\limits_{k_2\in\sigma} d^dk_2
\delta^d(p_1+p_2-k_1-k_2)\delta^d(p_1'+p_2'-k_1-k_2)
{u_4(p_1,p_2,k_1,k_2) u_4(k_1,k_2,p_1',p_2')\over v(k_1) v(k_2)}
\ee
This is $O(t^2)$ in general; but an exception occurs when $p_1+p_2=0$.
The integrations are then constraint by $\delta(k_1+k_2)$, and the graph 
becomes $O(t)$. This argument applies to any vertex of a graph, even if
it is a subgraph. Thus, in order for a one-loop graph to be $O(t)$
instead of $O(t^2)$, the total external momentum emerging from any one vertex 
must be zero.

Wegner and Houghton [11] sum the tree and one-loop graphs by means of a 
functional method, as follows. First, expand the action in powers of $f$:
\be
A[S+f]=A[S]+\sum_{k\in \sigma} P_k f_k 
+{1\over 2}\sum_{k\in\sigma} Q_k f_k f_{-k} +\cdots
\label{e:expand}
\ee 
where
\bea
P_k&=&\left[{\partial A[\phi]\over\partial f_k}\right]_{f=0}\no
Q_k&=&\left[{\partial^2 A[\phi]\over\partial f_k \partial f_{-k}}\right]_{f=0}
\label{e:PQ}
\eea
The terms represented by the dots in (\ref{e:expand}) may be omitted because
they do not contribute to $O(t)$. In the second term in 
(\ref{e:expand}),  we have a single $k$-sum instead of a sum
over two independent $k$'s, because of the restriction to a single loop 
momentum. This circumstance makes its possible to calculate the functional 
integral over $f$ to obtain
\be
Z=N\int DS e^{-\tilde A[S]}
\ee
where
\bea
\tilde A[S] &=&A[S]+t B[S]\no
B[S]&=&{1\over 2t}{\sum_{k\in\sigma}}
\left[\ln Q_k - {|P_k|^2\over Q_k }\right]
\label{e:bee}
\eea
The quantity $Q_k$ arises from one-loop graphs, while $|P(k)|^2$ arises
from tree graphs.

We now transform to the rescaled variables (\ref{e:kprime}) and 
(\ref{e:phiprime}). To first order in $t$, it is only necessary to do 
so in the first term of $\tilde A[S]$, since the 
second term is $O(t)$. We obtain, after a straightforward calculation,
\be
Z=N\int D\phi' e^{- A'[\phi']}
\ee
where
\bea
A'[\phi'] &=& A[\phi']+t \{B[\phi'] + C[\phi']\}\no
C[\phi]&=&{1\over t}\sum_{\alpha=2}^\infty
\sum_{|k_i|<1}
\delta(k) \left[\phi_{k_1}\cdots\phi_{k_\alpha}\right]
\left[ d+{\alpha\over 2}(2-\eta-d) 
- \sum_i k_i{\partial\over \partial k_i}
\right] u_\alpha(k)
\label{e:cee}
\eea
This is the result of Wegner and Houghton [11].

By expanding $B[\phi]$ and $C[\phi]$ in powers of $\phi$ we can express the
new action $A'[\phi]$ in the form (\ref{e:gaction}), and read off the new
coupling functions $u'_\alpha(k)$. The first-order change of the action 
can be written in the form
\be
A'[\phi]-A[\phi] = t\sum_{\alpha=2}^\infty\sum_{|k_i|<1}
\delta(k)\beta_\alpha(k) \phi(k_1)\cdots\phi(k_\alpha)
\ee
where
\be
\beta_\alpha(k)\equiv u'_\alpha(k)-u_\alpha(k)
\ee
Note that, to RG transformations, $u_\alpha(k)$ is a function of $t$ only, 
with  $\alpha$ and $k$ acting as labels for the type of coupling. 
Thus we can write
\be
{d u_\alpha(k)\over dt} = \beta_\alpha(k)
\label{e:wh}
\ee
which is an exact RG equation. The function $\beta_\alpha(k)$ depends on 
the $u_\alpha(k)$, but not on $t$ explicitly. This equation therefore
give the tangent vector to the trajectory at an arbitrary point 
in parameter space. Although this point
is identified as $t=0$ in the derivation, we can shift the origin of $t$
at will, because the equation is invariant under a translation in $t$.

Since the coupling function $u_\alpha$ obeys a differential equation in $t$, 
we can trace its evolution both forward and backward in $t$. This might
seem puzzling, since the RG transformation as defined appears to be 
irreversible. What renders it reversible is the fact that one and only 
one trajectory passes through any given point in the parameter space,
except at  a fixed point.

At this point, we can easily see that no derivative couplings are 
induced if none were present initially. Terms involving derivatives are 
generated by the momentum-dependent terms in $B[\phi]+C[\phi]$.
As we can see from (\ref{e:bee}) and (\ref{e:cee}), such terms can occur only
in $C[\phi]$, through the expression
\be
\sum_i k_i{\partial\over\partial k_i} u_\alpha(k)
\ee
If only non-derivative local couplings were present at the start, then the 
above vanishes except for $\alpha=2$, for which it gives a term proportional to $k^2$.
Therefore no derivative couplings are generated. This also shows that
a massless free field, which corresponds to the origin of the
parameter space, is invariant under RG. The origin is therefore a fixed 
point --- the Gaussian fixed point.
It can be seen that if there were no odd powers of the field initially, 
then none will be generated. The reason is that $Q_k$ in (\ref{e:bee}) is
even in the field. 

Graphs with $n$ external lines contribute to $u'_n$, and are  
shown in Fig.2 for $n=2,4,6$. In any one-loop graph, the $j$ 
external lines emerging from any vertex give rise to a factor 
$\int d^dx \phi^j(x)$, since they have total momentum zero. Thus,
a one-loop graph is generally proportional to a product of
such factors. For example, the graphs $a$, $b$, $c$ in Fig.2 lead 
to the following contributions to the action $A'[\phi]$:
\bea
G_a &=& u_8\int d^dx \phi^6(x)\no
G_b &=& {u_4u_6\over V}\int d^dx \phi^4(x)\int d^dy \phi^2(y)\no
G_c &=&  {u_4^3\over V^2}\left[\int d^dx \phi^2(x)\right]^3
\eea
The first contribution, coming from the ``diamond ring'' graph with
only one vertex, gives a local interaction. All others give uncorrelated
products of the fields, which correspond to non-local interactions of 
infinite range. The powers of the space-time volume $V$ in front 
of these expressions arise from the fact that the action should be $O(V)$.
All these uncorrelated non-local contribution are indeterminate in the
infinite-volume limit. The ambiguity clearly arises from the 
the infinitesimal RG step implemented with a sharp momentum cutoff. 
The products of field would have been correlated, if a gentle cutoff functions
had been used, or if the internal lines were integrated over a finite
instead of infinitesimal shell. However, the non-local terms are 
second order in the
bare couplings, and can be neglected in a 
linear approximation about the Gaussian fixed point.

The tree graph $d$ in Fig.2 contributes to $A'[\phi]$ a term of the form
\be
G_d = u_4^2\sum_{|k_i|<1}\delta(k_1+\cdots+k_6)\delta(|k_1+k_2+k_3|-1)
\phi_{k_1}\cdots\phi_{k_6}
\ee
which gives rise to a correlated non-local interaction. As shown in 
Ref.[11] this term gives rise to the ``non-trivial fixed point''
in $d=4-\epsilon\quad(\epsilon\to 0)$. But, since it is second order in 
the couplings, we shall ignore it here.

In view of the critical examination above, those results in Refs.[1]
and [12] pertaining to non-linear terms in the RG equation must be taken 
with reservation.

\section{Linearized RG Equations}

In the linear approximation, the action (\ref{e:action}) is closed under RG,
and we have a well-defined system.
To obtain the linearized RG equations, we need $B[\phi]$ defined in 
(\ref{e:bee}), in which the term $|P_k|^2$ can be neglected. A straightforward
calculation gives
\bea
Q_k &=& 1+r +\tilde Q\no
\tilde Q &=& \sum_{\alpha=2}^\infty \alpha (\alpha+1) V^{1-\alpha/2}
u_{\alpha+2}\sum_{|k_i|<1}\delta(k_1+\cdots+k_\alpha)\phi_{k_1}\cdots
\phi_{k_\alpha}
\eea
which is a sum over ``diamond rings,'' and is independent of $k$.
To first order in the $u_\alpha$, we have
\be
B[\phi]={1\over 2t}V_\sigma\tilde Q
\ee
where $V_{\sigma}$ is the volume of the thin momentum shell $\sigma$.

We quote the linearized RG equations generalized to an 
$N$-component field $\phi_i(x)\quad(i=1,\cdots,N)$ with $O(N)$ 
internal symmetry:
\bea
{du_{2n}\over dt} &=& (2n+d-nd) u_{2n} + S_d (n+1) (2n+N) u_{2n+2}\no
&&(n= 1,2,\ldots,\infty)
\label{e:linear}
\eea
where $S_d$ is the surface area of a unit  $d$-sphere divided by $(2\pi)^d$:
\bea
S_d &=& {2^{1-d}\pi^{-d/2}\over \Gamma(d/2)}\no
S_4 &=& {1\over 8\pi^2}
\eea

Let $\psi$ be the column matrix whose elements are $u_{2n}$. We can write
(\ref{e:linear}) in the form
\be
{d\psi\over dt}=M\psi
\ee
where $M$ is a matrix. Consider now the eigenvalue problem
\be
M\psi = \lambda\psi
\label{e:eigen}
\ee
The eigenvectors $\psi$ correspond to ``principal axes'' in the parameter 
space, along which we have the behavior $d\psi/dt = \lambda\psi$, or
\be
\psi(t) = \psi(t_0) e^{\lambda (t-t_0)}
\ee
The origin $t_0$ is arbitrary, except that it should be such that $\psi$
is small; but it should not correspond to the Gaussian fixed point, 
where $\psi\equiv 0$.

The eigenvalue $\lambda$ characterizes the trajectory tangent to the
corresponding principal axis at the Gaussian fixed point: 

(a) If $\lambda<0$, then $\psi\to 0$ as $t\to\infty$. The couplings constants
are said to be ``irrelevant.'' Under coarse-graining, they tend to 
the Gaussian fixed point, or triviality. On such a trajectory, the 
Gaussian fixed point is IR.

(c) If $\lambda>0$, then $\psi$ grows with $t$. The coupling constants 
are said to be ``relevant.'' Under coarse-graining, they tend to
go away from Gaussian fixed point. On such a trajectory the Gaussian
fixed point is UV, and the theory is nontrivial.
The trajectory is specified by some initial condition at 
an arbitrary point $t=t_0$, and it flows
away from the Gaussian fixed point. The latter can be reached by
letting $t\to\-\infty$, in which limit the couplings vanish. This is
asymptotic freedom.

(c) The case $\lambda=0$ corresponds to ``marginal'' coupling constants.
In this case, we have to go beyond the linear approximation in order
to determine the true behavior.

Using (\ref{e:linear}), we can put the eigenvalue equation 
(\ref{e:eigen}) in the form
\be
u_{2n+2} ={n(d-2)-d+\lambda\over S_d (n+1) (2n+N)} u_{2n}
\qquad (n=1,2,\cdots,\infty)
\ee
which is a recursion relation starting with $u_2=r/2$. To solve it in
terms of known functions, it is convenient to introduce a parameter
$a$ by writing the eigenvalue in the form
\be
\lambda = 2+(d-2)a
\ee
The recursion relation can then be put in the form
\be
u_{2n+2} ={(d-2) (a+n-1)\over 2S_d (n+1) (n+N/2)} u_{2n}
\ee
whose solution is
\be
u_{2n} ={r\over 2} \left({d-2\over 2S_d}\right)^{n-1} 
{a(a+1)\cdots (a+n-2)\over n! (n-1+N/2)(n-2+N/2)\cdots(1+N/2)}
\ee
The potential with these coupling constants is referred to as the 
``eigenpotential.'' Using the abbreviation
\be
z = {(d-2)\phi^2(x)\over 2S_d}
\ee
where $\phi^2=\sum_i\phi_i^2$, we have
\be
U_a(\phi^2(x))\equiv\sum_{n=1}^\infty u_{2n}\phi^{2n}(x)
= r{2S_d\over (a-1) (d-2)}\left[M(a-1, N/2, z)-1\right]
\ee
where $M(a,b,z)$ is the Kummer function [14]:
\be
M(a,b,z)=1+{a\over b}{z\over 1!}+{a(a+1)\over b(b+1)}{z^2\over 2!}+\cdots
={\Gamma(b)\over \Gamma(b-a)\Gamma(a)}
\int_0^1dt e^{zt} t^{a-1} (1-t)^{b-a-1}
\ee
If $a$ is a negative integer, the power-series breaks off to become a 
polynomial of degree $|a|$. Otherwise, its asymptotic behavior for 
large $z$ is given by
\be
M(a,b,z)\approx {\Gamma(b) z^{a-b}e^z\over \Gamma(a)}[1+O(z^{-1})]
\ee 
The eigenpotential $U_a(\phi^2)$ describes a field theory lying
on a trajectory tangent to a particular principal axis with
respect to the Gaussian fixed point. The principal axis is
identified only through the eigenvalue parameter $a$.

For a polynomial potential of even degree $2K$, then, we have $a=-2K$. 
The corresponding eigenvalues are
\be
\lambda=2[1-(d-2)K] \qquad (K=1,2,\ldots)
\ee
which is negative for $d=4$. In $d=3$ it is negative except for 
the marginal case of $K=1$; but that corresponds to a free theory.
Therefore, in $d>2$, all polynomial even potentials lead to triviality.

For $d=2$, the linear approximation breaks down completely. The reason 
is undoubtedly the formation of vortices that lead to the 
Kosterlitz-Thouless phase transition [15]. It would be very interesting
to discover vortices within the present framework, for in the existing 
literature they are simply put in by hand. We shall not
pursue this topic here, and will assume $d>2$ from now on.

\section{Non-Triviality and Asymptotic Freedom}

Nontrivial theories correspond to positive eigenvalues $\lambda>0$, which 
means that 
\be
a>-{2\over d-2}
\ee
They correspond to non-polynomial potentials with the following 
asymptotic behavior for large $\phi$:
\be
U(\phi^2)\sim\exp\left[{(d-2)\phi^2\over 2S_d}\right]
\ee
Nothing in canonical field theory rules out such a potential.

Sufficiently close to the Gaussian fixed point, the potential is
proportional to $r$, which evolves in $t$ according to
\be
r(t)=r(t_0)e^{\lambda (t-t_0)}=C e^{\lambda t}
\ee
with $C=r(t_0)\exp(-t_0)$. This is a running coupling constant, with 
a given renormalized value $r(t_0)$ at the reference point $t_0$. 
The theory is nontrivial, because the potential does not tend to
zero in the low-momentum limit. Instead, we have 
asymptotic freedom, corresponding to the fact that the potential 
vanishes in the limit $t\to -\infty$, which corresponds to infinite 
momentum.

In order to have spontaneous symmetry breaking on the semiclassical level,
the eigenpotential must have at least one minimum in $\phi$. The power
series expansion for the eigenpotential reads
\be
U_a(\phi^2)={4r S_d\over N(d-2)}\left[z + {az^2\over(1+N/2)2!} 
+ {a(a+1)z^3\over (1+N/2) (2+N/2) 3!} +\cdots\right]
\label{e:series}
\ee
A sufficient condition is that $U'(0)<0$, and $U>0$ for large $z$. 
The first is satisfied by choosing $r<0$. Asymptotically $U$ is 
proportional to $r[(a-1)\Gamma(a)]^{-1}$, the rest of the factors 
being positive. Thus we must have $(a-1)\Gamma(a)<0$, 
which is equivalent to $\Gamma(a-1)<0$. 
Using the formula $\Gamma(a)\Gamma(-a)=\pi/\sin(\pi a)$, and the fact that 
$\Gamma(a)$ is positive for $a>0$, we find that $a$ must be in one of 
the open intervals $(0,-1)$, $(-2,-3)$, {\it etc}. For a nontrivial theory, 
we have $\lambda>0$, or $2+(d-2)a >0$. Combining these requirements,
we obtain the sufficient condition
\be
-1<a<0
\ee
A family of eigenpotentials for this range of $a$,
and $d=N=4$, is plotted in Fig.3.

The eigenpotential $U_a$ corresponds to a theory that
lies on a trajectory tangent to a principal axis. Generally, we 
can consider a theory on an arbitrary trajectory, which is represented
near the Gaussian fixed point by a linear superposition of the
eigenpotentials. This gives us considerable freedom in choosing potentials.

The asymptotically free theory may be useful for models 
of the inflationary universe [16], for it offers a 
non-trivial quantum field theory with spontaneous symmetry breaking. 
From a philosophical  point of view, it seems more sensible to have a 
cosmological potential that was zero at the Big Bang and grow at 
decreasing energies, rather than the
conventional polynomial potential, which would have the opposite 
behavior if taken seriously. For such applications, one needs a
potential whose $\phi^2$ term is very small, of order $10^{-12}$
[17]. This turns out to be very natural in terms of our
eigenpotentials $U_a(\phi^2)$. As we can see from the power series
expansion (\ref{e:series}), the $\phi^2$ term is independent of $a$.
Therefore the difference of any two eigenpotentials
\be
V(\phi^2)=U_a(\phi^2)-U_{a'}(\phi^2) = {r(a-a')4 S_d\over N(1+N/2)(d-2)}
\left[ {z^2\over2!} 
+ {(a-a'+1)z^3\over (2+N/2) 3!} +\cdots\right]
\ee
has no $\phi^2$ term. Since this is a linear approximation, it means
that the $\phi^2$ term is $O(r^2)$. By taking $r<0$ and $a>a'$,
we make the potential go negative for small $\phi^2$. At large $\phi^2$ 
it must turn positive, because the the curves of the $U_a$ with 
different $a$'s intersect, as we can see in Fig.3. 
Therefore the potential has a negative minimum.

\section{Conclusion and Outlook}

We have shown that, near the Gaussian fixed point, all scalar theories
are trivial free fields in the low-energy limit, except for
a specific class with exponentially rising potentials, which are 
nontrivial at low energies, but become free in the high-energy limit.

The renormalized coupling constants used in this paper are not the same 
as the conventional ones in particle physics; the latter are
defined in terms of physical scattering amplitudes, which contain extra
momentum scales. The conventional renormalized coupling constants may be 
calculated by integrating the RG equations along a trajectory. 
We plan to address this topic in a separate paper.

The low-energy behavior of the asymptotically free theories lies
beyond the capability of the present formulation, because
the sharp momentum cutoff used here introduces ambiguities. It is
an important problem to implement Wilson's renormalization 
program with a gentle cutoff function, and extract results 
independent of the cutoff function.

An interesting extension of the present work would be to
make similar analyses of gauge fields and spinor fields. We hope 
the present paper will stimulate interest in this direction.

This work is supported in part by funds provided by the U.S. Department 
of Energy under cooperative agreement \# DE-FC02-94ER40818.
\newpage

\section*{References}
\begin{enumerate}
\item K. Halpern and K. Huang, Phys. Rev. Lett., {\bf 74}, 3526 (1995).
\item A.I. Larkin and D.E. Khumel'nitskii, Sov. Phys. JETP {\bf 29}, 1123
(1969).
\item K.G. Wilson, Phys. Rev. Lett. {\bf 28}, 248 (1972); 
K.G. Wilson and J.Kogut, Phys. Rep. {\bf C12}, 75 (1974).
\item B. Freedman, P. Smolinsky, and D. Weingarten, Phys. Lett.,
{\bf 113B}, 481 (1982).
\item  C.B. Lang, Nucl. Phys,, {\bf B265}, 630 (1986).
\item K. Huang, E. Manousakis, and J. Polonyi, Phys. Rev., {\bf 35}, 3187
(1987).
\item J. Kuti and Y. Shen, Phys. Rev. Lett., {\bf 60}, 85 (1988).
\item J. Kuti, L. Lin, and Y. Shen, Phys. Rev. Lett., {\bf 61}, 678 (1988).
\item R. Dashen and H. Neuberger, Phys. Rev. Lett., {\bf 50}, 1897 (1983).
\item J. Polchinsky, Nucl. Phys. {B231}, 269 (1984).
\item F.J. Wegner and A. Houghton, Phys. Rev., {\bf A8}, 401 (1973).
\item A. Hasenfratz and P. Hasenfratz, Nucl. Phys., {\bf B270}
[FS16], 687 (1986).
\item See, for example, K. Huang, {\it Quarks, Leptons, and Gauge Fields},
2nd Ed. (World Scientific, Singapore, 1992), Sec.9.5.
\item M. Abramowitz and I.A. Stegun, {\it Handbook of  
Mathematical Functions} (National Bureau of Standards, Washington, 1964),
p.503.
\item J. Kosterlitz and D. Thouless, J. Phys. C{\bf 6}, 1181 (1973).
\item A. Guth, Phys. Rev. {\bf D23}, 347 (1981);
A. Linde, Phys. Lett. {\bf 108 B}, 177 (1982). 
\item
A.A. Starobinsky, Phys. Lett. 117B, 175 (1982);
A.H. Guth and S.-Y. Pi, Phys. Rev. Lett. 49, 1110 (1982);
S.W. Hawking,  Phys. Lett. 115B, 295 (1982);
J.M. Bardeen, P.J. Steinhardt, and M.S. Turner, Phys. Rev.
     D28, 679 (1983).

\end{enumerate}
\newpage
\section*{Figure Captions}

Fig.1 A one-loop graph. The internal lines 
correspond to high-momentum components to be eliminated in 
the RG transformation. The external
lines represent low-momentum  components  left untouched.
\medskip\par\noindent
Fig.2 Contributions to renormalized $n$-field couplings for $n=2,4,6$.
\medskip\par\noindent
Fig.3  Eigenpotentials $U_a(\phi^2)$ as functions of 
$\phi\equiv\sqrt{\sum_{i=1}^N\phi_i^2}$, for $d=N=4$, in units
in which the momentum cutoff is unity. 
The ordinate is in arbitary units. From top to bottom, 
they correspond respectively  to  values of the 
the eigenvalue parameter $a$ uniformly spaced from $-0.999$ to $-0.001$. 
All of the potentials behave like $\exp\phi^2$ for large $\phi$, 
and lead to theories with asymptotic freedom. The limiting case $a=-1$ 
represents a $\phi^4$ potential, which gives a trivial theory.

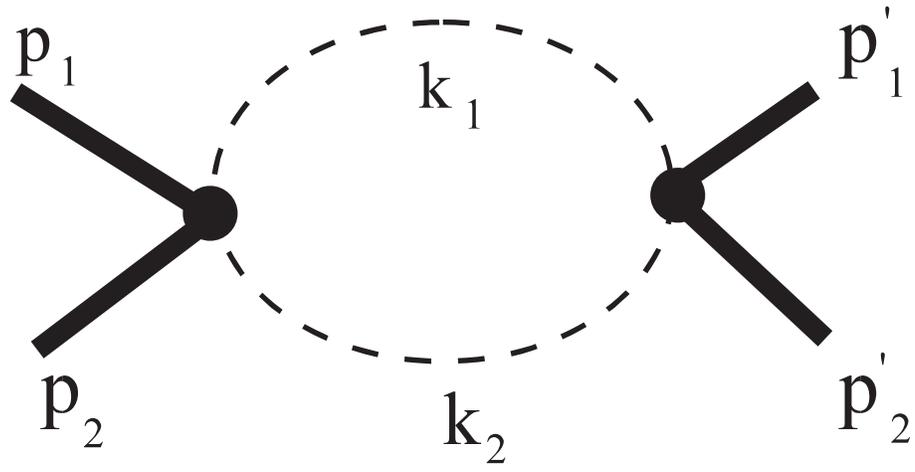
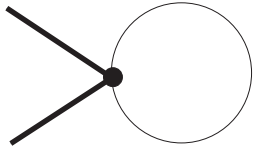
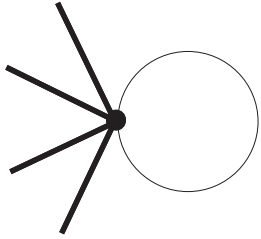
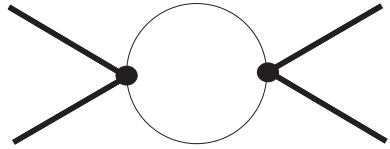
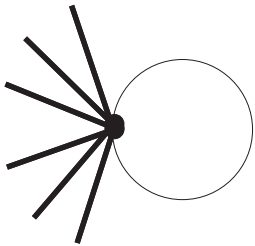
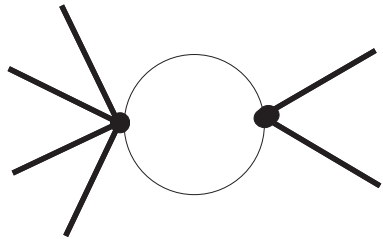
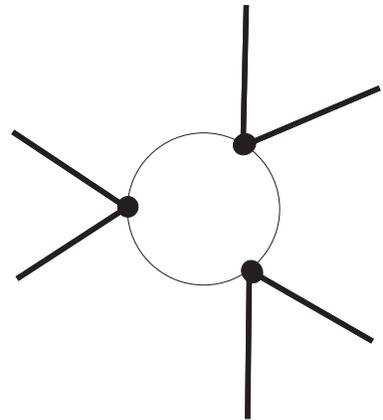
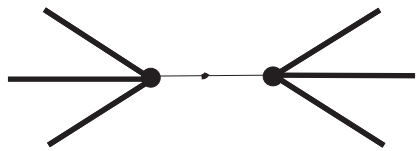
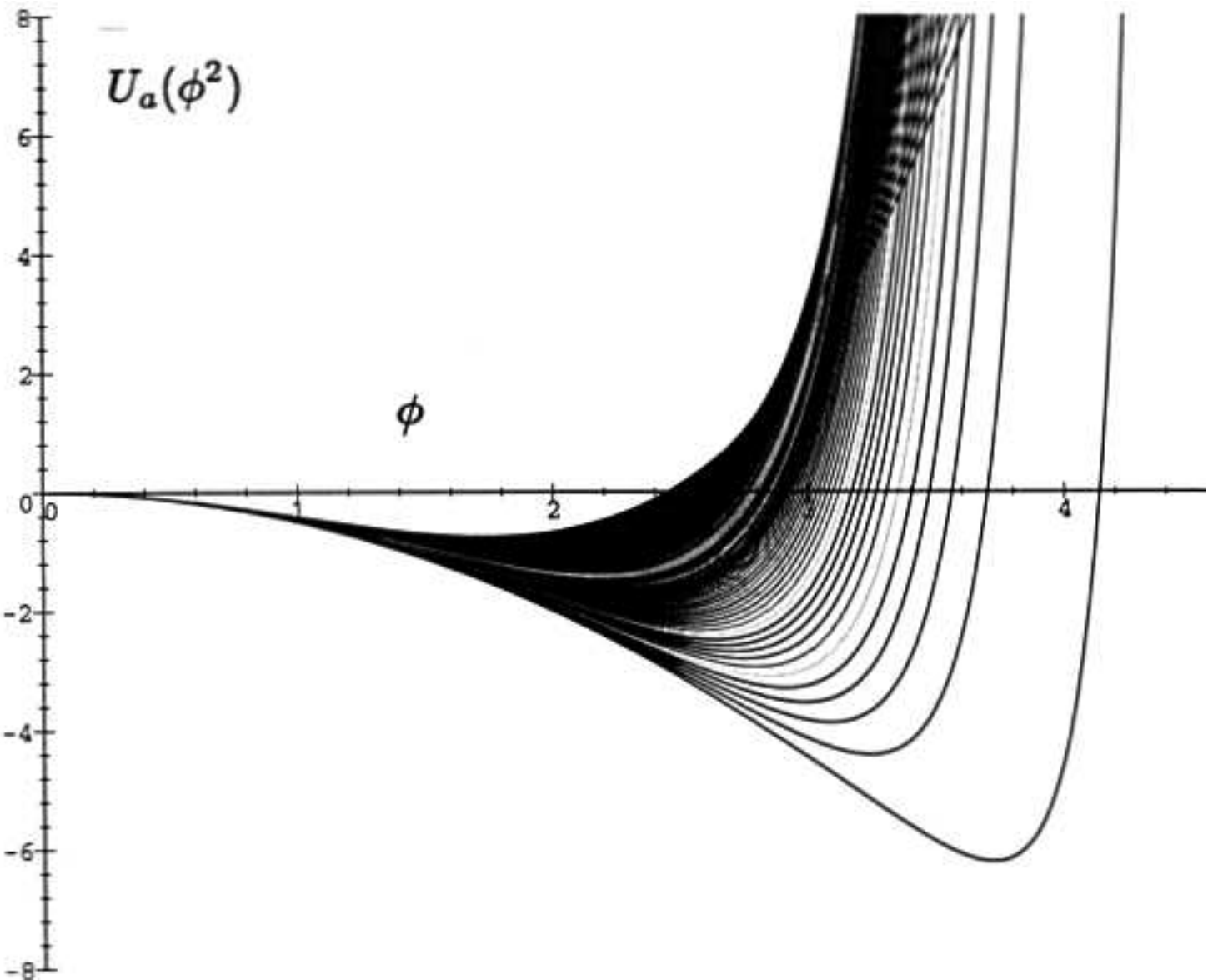
\end{document}